\documentclass[%
 reprint,
superscriptaddress,
 amsmath,amssymb,
 aps,
]{revtex4-2}

\usepackage{verbatim}
\usepackage{xcolor}
\usepackage{soul} 

\usepackage{color}
\usepackage{graphicx}
\usepackage{dcolumn}
\usepackage{bm}

\begin{document}

\preprint{APS/123-QED}

\title{The role of polarization field terms in a model for a cavity quantum material}

\author{Arwen Lloyd}
\affiliation{Department of Physics and Astronomy, The University of Manchester, Manchester M13 9PL, United Kingdom\\}%
\author{Adam Stokes}
\affiliation{School of Mathematics, Statistics and Physics, Newcastle University, Newcastle upon Tyne NE1 7RU, United Kingdom\\}
 \author{Alessandro Principi}
 \author{Ahsan Nazir}%
\affiliation{Department of Physics and Astronomy, The University of Manchester, Manchester M13 9PL, United Kingdom\\}%

\date{\today}

\begin{abstract}

Constructing models for cavity quantum materials requires a careful treatment of the light-matter coupling. In general, one must specify matrix elements constructed from the material wavefunctions, which are often unknown in a tight-binding framework. The Peierls substitution is frequently used to avoid introducing these additional parameters in the multi-center dipole (or Peierls) gauge, under the assumption that contributions from intraband and interband dipole moments can be neglected. We present the derivation of the Peierls gauge description, including these dipole moment terms, in the passive view of canonical transformations. We construct a toy model for a multi-band system with two sites, which we couple to a uniform field in the Coulomb, dipole, and Peierls gauges. We find that all polarization field terms are required to describe multi-band coupling in the Peierls gauge. The Peierls substitution can only be justified under restriction to a single band in one dimension, provided one also ignores self-polarization corrections arising from bands outside the retained subspace. However, these corrections are frequently non-negligible. More generally, the Coulomb, dipole, and Peierls gauges define distinct partitions of the composite system into the light and matter subsystems. We illustrate the implications of this subsystem relativity for observables such as the photon number and on the performance of orbital truncations in each gauge.
\end{abstract}

\maketitle

\section{\label{sec:intro}Introduction}

Advances in cavity quantum electrodynamics bring the possibility of extending the strong light-matter coupling regime from atomic and molecular systems to solids~\cite{schlawin_cavity_2022,garcia-vidal_manipulating_2021,bloch_strongly_2022}.
There has been growing interest in coupling the vacuum fields of a cavity to quantum materials, which opens the exciting possibility to control their properties.
Theoretical works propose mechanisms for inducing emergent many-body phenomena---such as superconductivity~\cite{schlawin_cavity-mediated_2019,chakraborty_long-range_2021} or quantum magnetism \cite{kiffner_manipulating_2019,chiocchetta_cavity-induced_2021}---and for tuning topological properties~\cite{dmytruk_controlling_2022,wang_cavity_2019} by coupling to cavity fields. Meanwhile, experiments have probed cavity control over metal-to-insulator transitions~\cite{jarc_cavity-mediated_2023}, and the enhancement of molecular superconductivity~\cite{thomas_exploring_2025} and of the fractional quantum hall effect in electron gases \cite{enkner_enhanced_2025}.
This combined interest has brought the need for, among other approaches, extending tight-binding models to describe coupling to the photonic fields~\cite{li_electromagnetic_2020}.

In general, describing the light-matter coupling requires direct knowledge of the matrix elements of the position or momentum operators. 
It is thus impossible to construct the light-matter interaction in a tight-binding framework, unless one is also provided with the underlying (Wannier or orbital) wavefunctions.
Furthermore, projecting the full theory into the low-energy sector to describe only a few electron bands coupled to the cavity field is expected to break the gauge invariance between truncated theories~\cite{li_electromagnetic_2020}.
This results from the freedom in defining the light and matter degrees of freedom, and in turn implies that the performance of truncated models depends on the gauge or canonical frame in which the truncation is performed~\cite{de_bernardis_breakdown_2018,stokes_gauge_2019,stokes_implications_2022,eyles_cauchy-schwarz_2025}.

It would be convenient to model the cavity system without explicit knowledge of the matrix elements associated with the light-matter coupling.
When an electron system is subject to a classical electromagnetic field, the Peierls substitution,
\begin{equation}
    t_{\ell\ell'}\to t_{\ell\ell'}\exp\left(iq\int_{\bm{R}_{\ell'}}^{\bm{R}_{\ell}}d\bm{s}\cdot\bm{A}_{T}(\bm{s})\right),
\end{equation}
can be made to remove minimal coupling terms in favor of a phase on the tight-binding hopping amplitudes \cite{luttinger_effect_1951,kohn_theory_1959,roth_theory_1962,blount_bloch_1962}.
This substitution has since been applied as a procedure to couple tight-binding models to electromagnetic fields in a cavity~\cite{dmytruk_controlling_2022,andolina_cavity_2019,mercurio_photon_2024,grunwald_cavity_2025},
and has sometimes been associated with the Coulomb gauge theory in this context~\cite{dmytruk_gauge_2021,passetti_cavity_2023,guerci_superradiant_2020,rao_non-fermi-liquid_2023}. 
However, for quantum fields of light, the transformation needed to bring about the Peierls phase factors mixes light and matter degrees of freedom. This enacts a canonical transformation, similar to that introduced in the multipolar gauge theory, and demands that the electromagnetic field terms are transformed to preserve the commutation relations~\cite{li_electromagnetic_2020}.  This introduces additional coupling via a polarization field in the full theory, which we will refer to as the Peierls gauge in this paper.

Our aim in this work is to introduce and study a toy model that allows us to assess the role of polarization-field terms in cavity quantum materials and thereby identify what is, and is not, captured by the Peierls substitution as an effective low-energy model. In this setting, we also clarify the distinction between Coulomb, dipole, and Peierls gauge descriptions of the light-matter system. We start by discussing the description of light-matter coupling in cavity quantum materials in Sec.~\ref{sec:theory}. We couple the continuum fields in the Coulomb gauge, which we transform to the Peierls gauge by projecting the matter fields onto a light-matter hybridized Wannier basis. We present a two-site toy model in which a single-electron is coupled to a uniform field in Sec.~\ref{sec:Two-site model}, which we can exactly diagonalize in the Coulomb, dipole, and Peierls gauges. Due to the anharmonicity of the model, we can couple separately to the intraband and interband dipole moment transitions by tuning the frequency of the field. We demonstrate the limitations of the Peierls substitution in these two cases, and we show how the neglected polarization field terms contribute to the Peierls gauge coupling in our model. We also discuss the gauge-relativity of the light and matter subspaces, including how this pertains to ground-state photon number predictions, and investigate the performance of orbital truncations in each gauge.

\section{\label{sec:theory}Light-matter coupling}

The quantization of the cavity fields begins with the solutions for Maxwell's equations in the absence of charges.
In the Coulomb gauge, we restrict to purely transverse electromagnetic fields, $\bm{\nabla}\cdot\bm{A}=0$.
The canonical operators may be decomposed onto a set of cavity eigenmodes as \cite{glauber_quantum_1991}
\begin{align}\label{Glauber_decomp}
    \hat{\bm{A}}_T(\bm{x}) &= \sum\nolimits_{\nu} \hat{Q}_{\nu} \bm{\phi}_{\nu}(\bm{x}), \\
    \hat{\bm{\Pi}}(\bm{x}) &= \sum\nolimits_{\nu}  \hat{\Pi}_{\nu} \bm{\phi}^{*}_{\nu}(\bm{x}),\notag
\end{align}
where $\bm{\phi}_{\nu}(\bm{x})$ denote the transverse field modes which solve the Maxwell equation of motion
\begin{equation}
    \omega_{\nu}^2 \,\bm{\phi}_{\nu} - \bm{\nabla} \times \left(\bm{\nabla} \times \bm{\phi}_{\nu}\right) = 0
\end{equation}
with frequency $\omega_{\nu}$, subject to the cavity boundary conditions. 
We have introduced the set of Glauber canonical variables which satisfy the commutation relations
\begin{align}\label{eqn:Glauber_comm}
    \left[\hat{Q}_{\nu},\hat{\Pi}_{\nu'}\right]&=i \delta_{\nu\nu'}, & \left[\hat{Q}_{\nu},\hat{\Pi}^{\dag}_{\nu'}\right]&=i U_{\nu\nu'},
\end{align}
with all other commutations of the canonical variables equal to zero. Here, $U_{\nu\nu'}$ is a unitary, symmetric matrix which maps field modes onto their complex conjugate as $ \bm{\phi}^{*}_{\nu}(\bm{x})=\sum\nolimits_{\nu'}U_{\nu\nu'}\bm{\phi}_{\nu'}(\bm{x})$.  As such, the Glauber variables can be expressed in terms of the boson operators as 
\begin{align}\label{eq:gmodes}
\hat{Q}_{\nu}&=A_{\nu}\left(\hat{a}_{\nu}+\sum\nolimits_{\nu'} U_{\nu\nu'}\hat{a}_{\nu'}^{\dag} \right), \\ \hat{\Pi}_{\nu}&= i\omega_{\nu}A_{\nu} \left(\hat{a}^{\dag}_{\nu}-\sum\nolimits_{\nu'} U^{*}_{\nu\nu'}\hat{a}_{\nu'}\right),\notag
\end{align}
with $A_{\nu}=(2\omega_{\nu})^{-1/2}$. 
 Since the matrix $U_{\nu\nu'}$ is entirely on-shell, $U_{\nu\nu'}\sim\delta(\omega_{\nu}-\omega_{\nu'})$, truncations on the photon modes should be performed on the frequency spectrum to preserve the Hermiticity of the gauge fields and the underlying bosonic algebra, $[\hat{a}_{\nu},\hat{a}_{\nu'}^{\dag}]=\delta_{\nu\nu'}$.

We can start by coupling the continuum light and matter fields in the Coulomb gauge. In the free, non-interacting theory, the electron field $\hat{\psi}(\bm{x})$ subject to a potential $V(\bm{x})$ is described by the Hamiltonian
\begin{equation}
\hat{H}_{\mathrm{m}}=\int d^3x\, \hat{\psi}^\dag(\bm{x}) \left(-\frac{\nabla^2} {2m} + V(\bm{x}) \right)\hat{\psi}(\bm{x}).
\end{equation}
When electrons are brought to interact with the electromagnetic fields under minimal coupling, the Hamiltonian in the Coulomb gauge is 
\begin{align}
    \hat{H}=&\int d^3x \, \hat{\psi}^\dag(\bm{x}) \left(\frac{(-i\bm{\nabla} - q\hat{\bm{A}}_T(\bm{x}))^2}{2m} + V(\bm{x}) \right) \hat{\psi}(\bm{x})\notag \\ 
    &+ \hat{U}_{\mathrm{ee}} + \hat{H}_{\mathrm{em}}.
\end{align}
The transverse electromagnetic fields, $\hat{\bm{E}}_{T}=-\hat{\bm{\Pi}}$ and $\hat{\bm{B}}=\bm{\nabla}\times\hat{\bm{A}}$, are purely photonic in this gauge, and their free dynamics is described by the Hamiltonian
\begin{equation}
\hat{H}_{\mathrm{em}}=\sum\nolimits_{\nu}\omega_{\nu}\left(\hat{a}^{\dag}_{\nu} \hat{a}_{\nu}+\frac{1}{2}\right).
\end{equation}
The longitudinal electric field, however, has been removed using
Gauss' law, $\bm{\nabla}\cdot\hat{\bm{E}}=q\hat{\psi}^{\dag}(\bm{x})\hat{\psi}(\bm{x})$, 
and gives rise to the 
Coulomb interaction term between electrons, 
\begin{equation}
\hat{U}_{ee} = \frac{1}{2} \int d^3x \,\hat{\bm{E}}_L(\bm{x})^2.
\end{equation}
In the absence of the cavity, this interaction is conventionally normal-ordered with respect to the vacuum~\cite{cohen-tannoudji_photons_1997,altland_condensed_2023}.
However, it should be emphasized that this interaction depends on the boundary conditions imposed by the cavity on the total electric field, ${\hat{\bm E}}$. In general, an image charge distribution must be introduced to generate the required longitudinal part, ${\hat{\bm E}}_L$, in the Coulomb gauge~\cite{power_quantum_1982}.

In order to construct a tight-binding model, we would like to decompose the electron field onto a basis which is sufficiently localized at the atomic scale. 
We can assume that electrons are subject to a periodic lattice potential, $V(\bm{x})$, and that we can construct an orthonormal basis of Wannier functions, $w_{\ell}(\bm{x})$. The electron field is decomposed as
\begin{equation}\label{eqn:Wannier}
    \hat{\psi}(\bm{x})=\sum\nolimits_{\ell}w_{\ell}(\bm{x})\hat{c}_{\ell},
\end{equation}
where $\ell=(\bm{R},\alpha)$ labels the Wannier orbital at site $\bm{R} $ in band $\alpha$, and $\hat{c}_{\ell}$ and $\hat{c}_{\ell}^{\dag}$ describe the fermion canonical algebra by $\{\hat{c}_{\ell},\hat{c}_{\ell'}^{\dag}\}=\delta_{\ell\ell'}$.
We assume that the field $\hat{\psi}(\bm{x})$ can be projected onto a set of bands whose Wannier functions are well localized on the scale of the lattice. In practice, constructing maximally-localized Wannier functions may require mixing the band index~\cite{marzari_maximally_2012}.  

Expressing the fields in the Glauber and Wannier bases, we can determine the form of the coupling which we must introduce to the tight-binding model in the Coulomb gauge.  Before projection to a low-energy subspace of interest, the Hamiltonian is written in full as
\begin{equation}
    \hat{H}=\sum\nolimits_{\ell\ell'} t_{\ell\ell'}\hat{c}_{\ell}^{\dag} \hat{c}_{\ell'} - \sum\nolimits_{\nu}\hat{J}^{\dag}_{\mathrm{p},\nu} \hat{Q}_{\nu} + \hat{H}_{\mathrm{dia}}+\hat{U}_{\mathrm{ee}}+ \hat{H}_{\mathrm{em}}
\end{equation}
where $t_{\ell\ell'}$ are the hopping amplitudes, 
\begin{equation}
    \hat{J}_{\mathrm{p},\nu}=-\frac{iq}{2m}\int d^3x\,\bm{\phi}^{*}_{\nu}(\bm{x})\!\cdot\!\left[(\bm\nabla \hat{\psi}^{\dag})\hat{\psi}(\bm{x})-\hat{\psi}^{\dag}(\bm{x})(\bm\nabla \hat{\psi})\right]
\end{equation}
are field mode components of the paramagnetic current operator, and the diamagnetic current term can likewise be decomposed as
\begin{equation}
\hat{H}_{\mathrm{dia}}=-\frac{q^2}{2m}\sum\nolimits_{\nu\nu'} \hat{Q}^{\dag}_{\nu}\hat{Q}_{\nu'}{\int d^3x\,\bm{\phi}^{*}_{\nu}(\bm{x})\!\cdot \!\bm{\phi}_{\nu'}(\bm{x})\hat{\psi}^{\dag}(\bm{x})\hat{\psi}(\bm{x})}.
\end{equation}

For localized electron subsystems, such as atoms or molecules, the Power-Zienau-Wooley (PZW) transformation is often used to transform the Coulomb gauge theory to the multipolar gauge \cite{cohen-tannoudji_photons_1997}. An electric dipole approximation (EDA), $\lambda\gg L$ (where $L$ is the size of the matter subsystem), is typically made to remove higher-order multipolar interaction terms, resulting in the well-known dipole gauge Hamiltonian.
This gauge transformation has since been generalized to periodic systems by performing a site-dependent canonical transformation on the electronic operators~\cite{li_electromagnetic_2020}.  The transformation extends the Peierls subsitution to quantized cavity fields and the resulting Hamiltonian is known as the multi-center dipole gauge. For the purpose of clarity, we refer to this canonical frame as the Peierls gauge throughout this paper to distinguish it from the dipole gauge picture. The details of the transformation are given in the passive view in Appendix~\ref{sec:PZW transformation}, which we will also summarise below.

We begin the transformation by decomposing the matter field onto hybridized Wannier orbitals,
\begin{align}\label{eqn:mPZW}
    \hat{\psi}(\bm{x})
    &=\sum\nolimits_{\ell} w_{\ell}(\bm{x})e^{-iq\hat{\chi}_{\ell}(\bm{x})} \hat{\bar{c}}_{\ell},
\end{align}
where the transformed site operators, $\hat{\bar{c}}^{}_{\ell}$ and $\hat{\bar{c}}^{\dag}_{\ell}$, form a new set of fermion operators. The photonic phases, $\hat{\chi}_{\ell}(\bm{x})=-\int_{\bm{R}_{\ell}}^{\bm{x}}d\bm{s}\cdot\hat{\bm{A}}_{T}(\bm{s})$, are chosen as line integrals centered on the lattice sites of each orbital $\ell$. 
This transformation is analogous to what has been done in Refs.~\cite{luttinger_effect_1951,kohn_theory_1959,roth_theory_1962,blount_bloch_1962} for classical electromagnetic fields. 
The key difference is that transforming the orbitals by a quantized photonic phase enacts a canonical transformation, which mixes the photon and electron subspaces between the two frames.

The photon operators must also transform to complete the transformation on the composite system. This will introduce a polarization field to the light-matter description. 
While the transverse magnetic potential $\hat{\bm{A}}_{T}$ is invariant, its conjugate field transforms as
\begin{equation}\label{eq:ph_trans}
    \hat{\bar{\bm{\Pi}}}(\bm{x})=\hat{\bm{\Pi}}(\bm{x})-\hat{\bm{P}}_{T}(\bm{x})
\end{equation}
where $\hat{\bm{P}}_{T}(\bm{x})=\sum_{\nu} \hat{P}_{\nu}\bm{\phi}^{*}_{\nu}(\bm{x})$ is a transverse polarization field, whose explicit expression is given in Eq.~\eqref{eqn:multi-PT}. The operators $\hat{P}_{\nu}$ are chosen to ensure that the photonic variables $\hat{\bar{\Pi}}_{\nu} =\int d^3x\,\bm{\phi}_{\nu}(\bm{x})\cdot\hat{\bar{\bm{\Pi}}}(\bm{x})$ and $\hat{\bar{Q}}_{\nu}\equiv\hat{Q}_{\nu}$ satisfy the Glauber commutation relations in Eq.~\eqref{eqn:Glauber_comm}, and commute with the set of transformed fermion operators.

A local EDA is also made in the multi-center PZW transformation by placing a truncation,  $\nu\in\Lambda$,  on the photon modes. We assume that the dominant cavity modes vary slowly in space with respect to the lattice scale. 
Under this assumption, the transformation removes the transverse magnetic potential from the minimal coupling, up to small magnetic corrections which can be neglected. In its place, the hopping amplitudes $t_{\ell\ell'}$ are dressed by the photonic Peierls phases, 
\begin{equation}\label{eq:peierls_phases}
    \hat{\chi}_{\ell\ell'}=\int_{\bm{R}_{\ell'}}^{\bm{R}_{\ell}}d\bm{s}\cdot\hat{\bm{A}}_{T}(\bm{s}).
\end{equation}
The polarization field, which was introduced in Eq.~\eqref{eq:ph_trans}, will also couple the cavity photons directly to dipole moments of the electron subsystem. 
In the local EDA, the transverse modes of this field are given by 
    \begin{equation}
        \hat{P}_{\nu}= q \sum\nolimits_{\ell\ell'} \bm{D}'_{\ell\ell'}\cdot\bm{\phi}_{\nu}(\bm{R}_{\ell'})e^{iq\hat{\chi}_{\ell\ell'}}\hat{\bar{c}}^{\dag}_{\ell} \hat{\bar{c}}_{\ell'}
    \end{equation}
where $\bm{D}'_{\ell\ell'}=\int d^3x \,w^*_{\ell}(\bm{x}) (\bm{x}-\bm{R}_{\ell'})w_{\ell'}(\bm{x})$ is the centered dipole moment between Wannier orbitals $\ell$ and $\ell'$. 

The full light-matter coupling is described in the Peierls gauge by the Hamiltonian
\begin{align} \label{eq:H_peierls_quantum}
    \hat{H}=&\sum\nolimits_{\ell\ell'} t_{\ell\ell'}e^{iq\hat{\chi}_{\ell\ell'} }\hat{\bar{c}}^{\dag}_{\ell} \hat{\bar{c}}_{\ell'} + \frac{1}{2} \sum\nolimits_{\nu} \left(\hat{P}^{\dag}_{\nu}\hat{\bar{\Pi}}_{\nu}+\hat{\bar{\Pi}}^{\dag}_{\nu}\hat{P}_{\nu}\right) \notag \\
    &+ \frac{1}{2}\sum\nolimits_{\nu}\hat{P}^{\dag}_{\nu} \hat{P}_{\nu} + \hat{U}_{\mathrm{ee}} + \hat{\bar{H}}_{\mathrm{em}} 
    \end{align}
where we identify the first term as the Peierls substitution on the free electron Hamiltonian. The second and third terms on the right-hand side of Eq.~\eqref{eq:H_peierls_quantum} describe a dipolar interaction and the self-interaction of the polarization field respectively, and arise from the transformation on the photonic sector. The free photon Hamiltonian is now identified as
$\hat{\bar{H}}_{\mathrm{em}}=\sum\nolimits_{\nu} \omega_{\nu}(\hat{\bar{a}}^{\dag}_{\nu} \hat{\bar{a}}_{\nu}+1/2)$,
where $\hat{\bar{a}}_{\nu}$ and $\hat{\bar{a}}^{\dag}_{\nu}$ are defined as before in Eq.~\eqref{eq:gmodes} for the transformed Glauber modes, $\hat{\bar{Q}}_{\nu}$ and $\hat{\bar{\Pi}}_{\nu}$. We leave the discussion of the transformed Coulomb interaction term,  $\hat{U}_{\mathrm{ee}}$, to the Appendix. 

Truncating the photon frequency spectrum will not violate unitary invariance between the Coulomb and Peierls gauges, provided one truncates the modes of the polarization field also~\cite{cohen-tannoudji_photons_1997}. It is possible to define a unitary transformation connecting the photon mode-truncated theories, which we discuss in Appendix~\ref{sec:PZW transformation}. 
Nonetheless, the transformation does mix electronic states between bands. Restricting the matter field operator to only a few bands, $\hat{\psi}(\bm{x})=\sum_{\bm{R},\alpha\in\Lambda_{m}}w_{\bm{R},\alpha}(\bm{x})\hat{c}_{\bm{R},\alpha}$,  
will often lead to a well-known breakdown of gauge-invariance between the truncated Coulomb and dipole or Peierls gauge models.

Before we conclude this section, we highlight that a useful feature of 1D electron systems is that the intraband dipole moments, $D'_{R\alpha,R'\alpha}$, can be made to vanish by exploiting a  gauge freedom of the electronic wave functions. This is because in 1D the Wannier functions can always be chosen as real, symmetric or antisymmetric, and exponentially localized for a symmetric potential $V(x)$~\cite{kohn_analytic_1959}. This simplifies the electromagnetic coupling to 1D single-band models, since the polarization field terms vanish from the low-energy theory, up to the inclusion of interband terms in the self-polarization, as we will see in Sec.~\ref{sec:pol}.

\section{\label{sec:Two-site model}Toy model }

\subsection{Electron in a 1D potential}

The multi-center dipole gauge, or Peierls gauge, can also be defined for molecular systems, for which we can construct a basis of atomic-like orbitals. A single-electron model can be solved by numerical diagonalization, and so we can readily investigate the role of the polarization field terms. By choosing a highly anharmonic potential, we can form a basis with ``bands'' of localized orbitals, similar to the Wannier functions of extended systems (and, as we will see, with similar symmetries on the dipole moment operators).

\begin{figure}[t]
    \includegraphics[width=\linewidth]{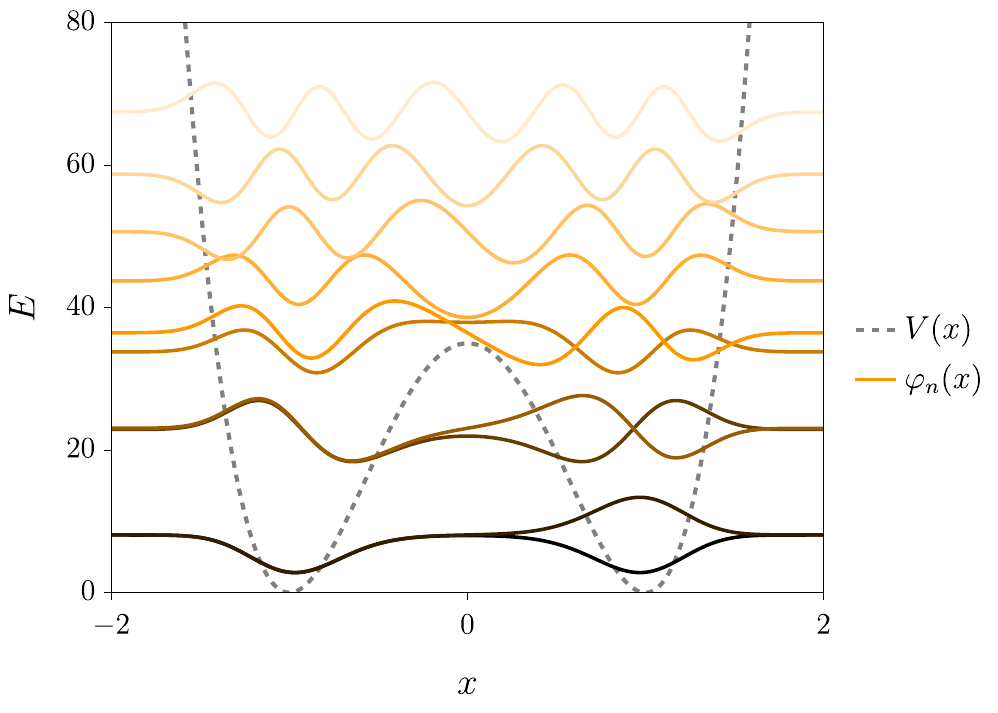}
\caption{The double-well potential $V(x)$ is drawn for $U=35$ and $b=1$. The wavefunctions, $\varphi_n(x)$, of the first ten eigenstates are shown (not to scale), each zeroed on their eigenenergy $\varepsilon_n$, for mass $m=1$. }
\label{fig:dipole}
\end{figure}

A double-well potential of the form
\begin{equation}\label{eq:dipole_pot}
    V(x)=\frac{U}{b^4}\left(x^2-b^2\right)^2
\end{equation}
generates a single-particle spectrum with a number $m$ of pairs of nearly degenerate states. These states are localized within the two wells, with even and odd symmetry about its center, as it is shown for a choice of parameters in Fig.~\ref{fig:dipole}. The double-well potential is often employed to verify the accuracy of truncations in the context of the quantum Rabi model, which has been a source of discussion on the breakdown of gauge invariance of truncated models in the Coulomb and dipole gauges \cite{de_bernardis_breakdown_2018,stokes_gauge_2019,di_stefano_resolution_2019,savasta_gauge_2021, stokes_gauge_2024,eyles_cauchy-schwarz_2025}.

In order to define a site-dependent canonical transformation, we will first introduce a new electronic basis. We rotate the bound wavefunctions, $\varphi_{n\leq 2m}(x)$, as follows:
\begin{equation}
    \psi_{i=1,2;\alpha}(x)=\frac{1}{\sqrt{2}}(\varphi_{2\alpha-1}(x)\pm\varphi_{2\alpha}(x))
    \end{equation}
where $\alpha=1,\ldots,m$ indicates the bound pair and $i=1,2$ will label one of the two wells. In this new basis, the states in each nearly-degenerate pair have been localized in mutual wells of the potential, and the wave functions $\psi_{i,\alpha}$ are analogous to Wannier functions or atomic orbitals for this two-site model.
We can restrict the Hamiltonian to the single-electron subspace without breaking gauge invariance, since the relevant operators remain one-body on the electron subspace following the transformation. For notational clarity, we denote states in the electronic eigenbasis by $|n\rangle$ and the rotated, Wannier-like basis by $|i,\alpha\rangle$ (with $|2\alpha-1\rangle=|1,\alpha\rangle$ and $|2\alpha\rangle=|2,\alpha\rangle$ for $\alpha>m$).

To couple the double well to a uniform cavity field, we can assume that the electron field has been restricted to the $x$-axis as $\psi(\bm{x})\sim\psi(x)\delta(y)\delta(z)$. We introduce a mode function, $\bm{\phi}_0(\bm{x})$, which is locally uniform along the $x$-axis. The electronic system will appear to couple to one-dimensional, uniform photonic fields once we integrate over the $y,z$ dimensions of the cavity. By fixing the magnitude of the field mode as $\phi_0(x)= \bm{e}_{x}\cdot\bm{\phi}_0(\bm{x}) |_{y,z=0}=\sqrt{2\omega} A$, the photonic fields are locally described by $\hat{A}(x)=A\left(\hat{a}+\hat{a}^\dag\right)$ and $\hat{\Pi}(x)=-i\omega A \left(\hat{a}-\hat{a}^\dag\right)$ in the $x$-direction.

In the Coulomb gauge, the Hamiltonian for the single electron coupled to the uniform field is given by 
\begin{equation}\label{hC}
        \hat{H}=\hat{H}_\mathrm{m} - \hat{J}_{\mathrm{p}} A \left(\hat{a}+\hat{a}^{\dag}\right) + \frac{q^2}{2m}  A^2 \left(\hat{a}+\hat{a}^{\dag}\right)^2 + \hat{H}_{\mathrm{em}}.\!\!
\end{equation}
The free electron and photon Hamiltonians are given by $\hat{H}_\mathrm{m}=\sum_{n}\varepsilon_n|n\rangle\langle n|$ and $\hat{H}_{\mathrm{em}}=\omega(\hat{a}^{\dag}\hat{a}+1/2)$,  
and $\hat{J}_{\mathrm{p}}=\sum_{jk} J_{jk}|j\rangle \langle k|$ is the integral of the paramagnetic current operator, whose matrix elements are given by
\begin{equation}
J_{jk}=-\frac{iq}{2m} \int dx\,\left(\frac{\partial \varphi^*_{j}(x)}{\partial x} \varphi_{k}(x)-\varphi^*_{j}(x) \frac{\partial\varphi_{k}(x)}{\partial x}\right),
\end{equation} 
or more commonly in terms of the momentum operator as $J_{jk}=q\langle j|\hat{p}|k\rangle/m$.

We can perform the PZW transformation about the origin to introduce the dipole gauge Hamiltonian,
\begin{equation}\label{hd}
    \hat{H}= \hat{H}_\mathrm{m} - i \omega q \hat{D} A \left(\hat{a}-\hat{a} ^{\dag}\right)+  \omega A^2 q^2  \hat{D}^2 + \hat{H}_{\mathrm{em}},
\end{equation}
where $\hat{a}$ and $|j\rangle$ are now defined in the dipole gauge.
The transverse polarization field is described by a uniform mode contribution, $\hat{P}_0 =\sqrt{2\omega} A q \hat{D}$, 
 and the dipole moment operator $\hat{D}=\sum_{jk}D_{jk}|j\rangle \langle{k}|$ is simply given by the position matrix elements via $D_{jk}=\int dx\,\varphi^*_{j}(x)x\varphi_{k}(x)=\langle j|\hat{x}|k\rangle $.

In the Peierls gauge, the matter matrix elements are dressed by Peierls factors in the rotated electronic basis. The Hamiltonian can be written as
\begin{align}\label{hP}
    \hat{H}=& \sum\nolimits_{ij,\alpha\beta} h_{i\alpha,j\beta} e^{i q (R_{i,\alpha}-R_{j,\beta}) \hat{A}} |i,\alpha\rangle\langle j,\beta|\\& -i \omega A q  \left(\hat{D'}\hat{a}-\hat{a}^{\dag}\hat{D'}\right) +\omega A^2 q^2 \hat{D}'^2+ \hat{H}_{\mathrm{em}}\notag
\end{align}
where $h_{i\alpha,j\beta}$ are the matrix elements of the free electron Hamiltonian, $\hat{H}_\mathrm{m}$, and $\hat{a}$ and $|i,\alpha\rangle$ are defined in the Peierls gauge.
The uniform polarization field mode, $\hat{P}'_0=\sqrt{2\omega} Aq\hat{D}'$, has introduced a hybridized dipole-moment operator,
\begin{equation}
\hat{D}'=\sum\nolimits_{ij,\alpha\beta}D'_{i\alpha,j\beta} e^{i q (R_{i,\alpha}-R_{j,\beta}) \hat{A}} |i,\alpha\rangle\langle j,\beta|,
\end{equation}
to the description of the light-matter interaction, where $D'_{i\alpha,j\beta}=\int dx\,\psi^*_{i,\alpha}(x)(x-R_{j,\beta})\psi_{j,\beta}(x)$ are the matrix elements of the site-centered dipole moment operator.
In the dipole gauge basis, we can see that the two dipole moments are related by $\hat{D}'=\sum_{jk}D'_{jk}|j\rangle \langle k|_{\mathrm{dipole}}=\hat{D}-\hat{d}$, where $\hat{d}=\sum_{i,\alpha} R_{i,\alpha}|i,\alpha\rangle\langle i,\alpha|$ is a purely on-site dipole moment operator.

\begin{figure}[t]
\includegraphics[width=\linewidth]{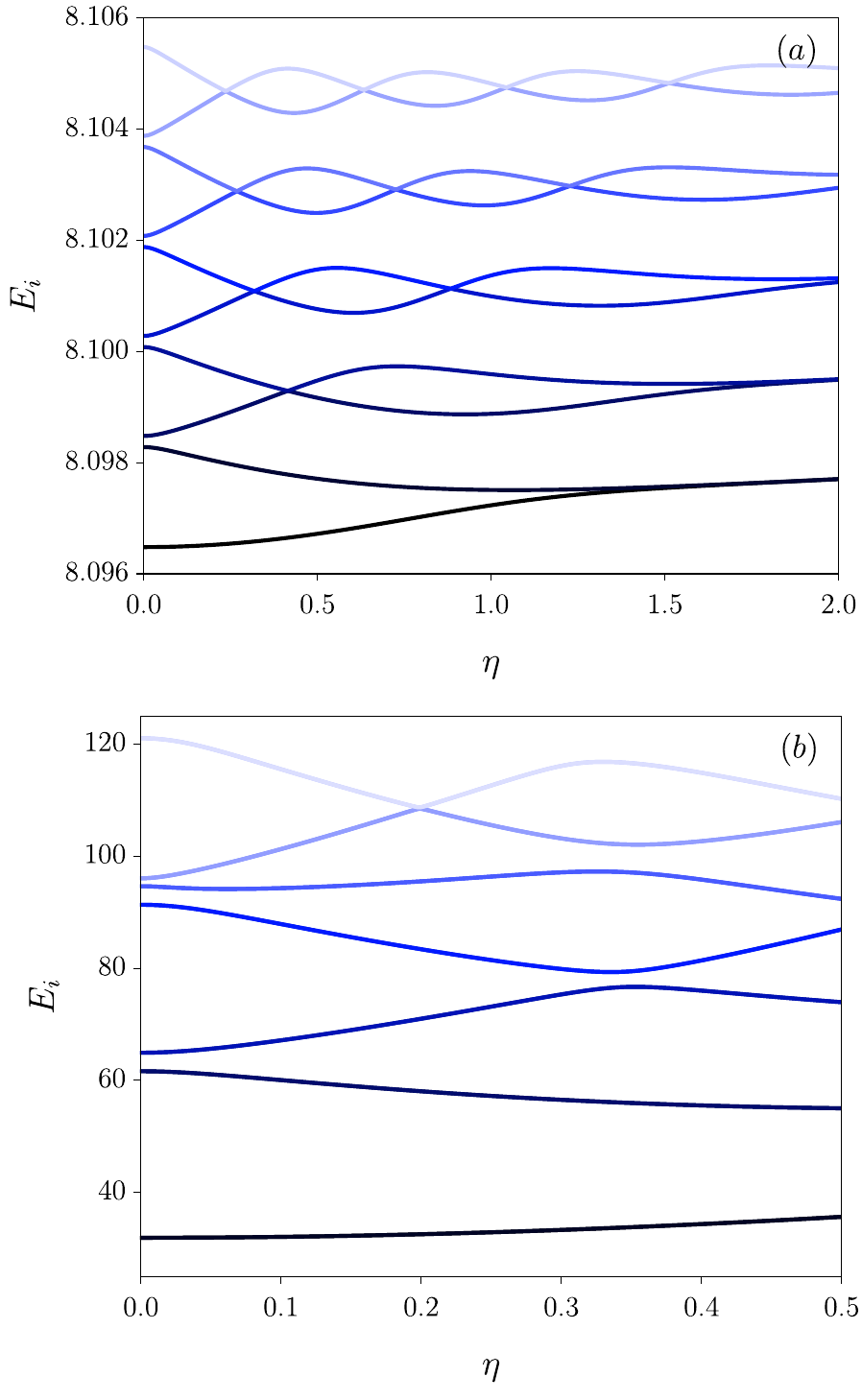}
\caption{(a) The energy spectrum for the first 10 eigenstates of the model is plotted for $\omega=0.9 \omega_{12},U=35$, and $b=1$ as a function of the coupling strength $\eta$. (b) Likewise, the spectrum for the first 14 eigenstates is plotted for $\omega=0.9 \omega_{13},U=150$, and $b=1$ (each line shown is nearly doubly degenerate). The electron charge and mass are taken as $q=1$ and $m=1$ in both cases.} 
\label{fig:spectrum}
\end{figure}

While we cannot grasp the periodicity of extended systems in this two-site toy model, we capture the same symmetries on the electric dipole moment in the Peierls gauge.
Firstly, the dipole moment operator will approximately vanish for transitions within each bound pair, $D^{(\prime)}_{1\alpha,2\alpha}=\int dx\, \psi_{1,\alpha}^*(x) x \psi_{2,\alpha}(x)\approx0$, as is the case for intraband terms in 1D systems. Secondly, by choosing to perform the multi-center PZW transformation about the orbital centers, $R_{i,\alpha}=\int dx\,|\psi_{i,\alpha}(x)|^2 x$, the on-site dipole moments have vanished in the Peierls gauge by definition.
As such, only the ``interband'' dipole moment terms will enter the Hamiltonian in Eq.~\eqref{hP}.

We are interested in photon frequencies which are close to resonance with the first two orbital transitions. We consider the frequencies $\omega=(1-\delta)\omega_{1n}$ for $n=2,3$, where $\omega_{1n}=\varepsilon_{n}-\varepsilon_{1}$ and \mbox{$\delta=0.1 $} is the fractional detuning. When $\omega\sim\omega_{12}$, the photon field couples resonantly to transitions between nearly-degenerate states, and is analogous to ``intraband'' transitions in extended cavity systems. Likewise, when $\omega\sim\omega_{13}$, the photon field will couple to ``interband'' transitions between the bound pairs. 
Since $\omega_{12}\ll\omega_{13}$, we can explore these two coupling regimes separately, as would be the case for an electron system with flat bands.

We produce the cavity spectra for both of these regimes in Fig.~\ref{fig:spectrum}, which converges in all gauges for sufficiently high truncation cutoffs on the orbital and photon number bases. We parameterize the strength of the light-matter coupling by the dominant transition dipole moment for each frequency. For $\omega\sim\omega_{12}$, the dimensionless coupling parameter is given by $\eta=q A |D_{12}|$, where $\eta\sim1$ describes a strong coupling regime. We note that for $\omega\sim\omega_{13}$ the dominant transition is $D_{14}$ (or $D_{23}$ equivalently) due to the symmetry of the orbitals, leading to the coupling parameter $\eta= q A |D_{14}|$.

\subsection{Gauge-relativity of light and matter}\label{sec:photon}

Canonical transformations mix the cavity system's electron and photon degrees of freedom.  The separation of ``light'' and ``matter'' in the composite system is  gauge relative, since these subsystems are associated with different physical observables in each gauge or canonical frame~\cite{stokes_implications_2022,stokes_gauge-relativity_2023}. 
When seen in the active view, a canonical transformation is enacted by a unitary transformation on vectors and operators; a physical state or observable is therefore associated with a different vector or operator in each frame~\cite{cohen-tannoudji_photons_1997}.
In the passive view, which we take throughout this paper, the changes to the partition of the Hilbert space are more explicit. The electron and photon canonical operators are transformed between frames $g$ and $g'$ by $
    \hat{a}_{\nu}|_g = \mathcal{U} \hat{a}_{\nu}|_{g'} \mathcal{U}^{\dag}$ and $
    \hat{c}_{\ell}|_{g} = \mathcal{U} \hat{c}_{\ell}|_{g'}\mathcal{U}^{\dag}$. While vectors and operators are not transformed directly in this view, they admit new representations when expressed in terms of the new set of canonical operators.

Expectation values for physical observables must be invariant under transformation. 
For example, the transverse current can be determined from Maxwell's equations in the Coulomb, dipole, or Peierls gauge~\cite{li_electromagnetic_2020}. The associated operator, $\hat{J}_{T,0}$ (for the uniform mode), is the same in each gauge in the passive view, but finds a unique representation in terms of the canonical operators:
\begin{equation}
    \hat{J}_{T,0}= i\left[ \hat{H}^{(g)}_{\mathrm{int}},\hat{\Pi}_0|_g\right]=-\frac{\delta \hat{H}^{(g)}_{\mathrm{int}}}{\delta \hat{Q}_0}+\frac{\partial}{\partial t} \hat{P}_0^{(g)} 
\end{equation}
where $\hat{H}_{\mathrm{int}}^{(g)}=\hat{H}-\hat{H}_{\mathrm{em}}^{(g)}$,  with $\hat{H}_{\mathrm{em}}^{(g)}=\omega(\hat{a}^{\dag}\hat{a}|_g+1/2)$, and $\hat{P}_{0}^{(g)}$ is the polarization field mode which is introduced in the gauge $g$, with $\hat{P}_0^{(\mathrm{Coulomb})}=0$.

\begin{figure}[t]
    \includegraphics[width=\linewidth]{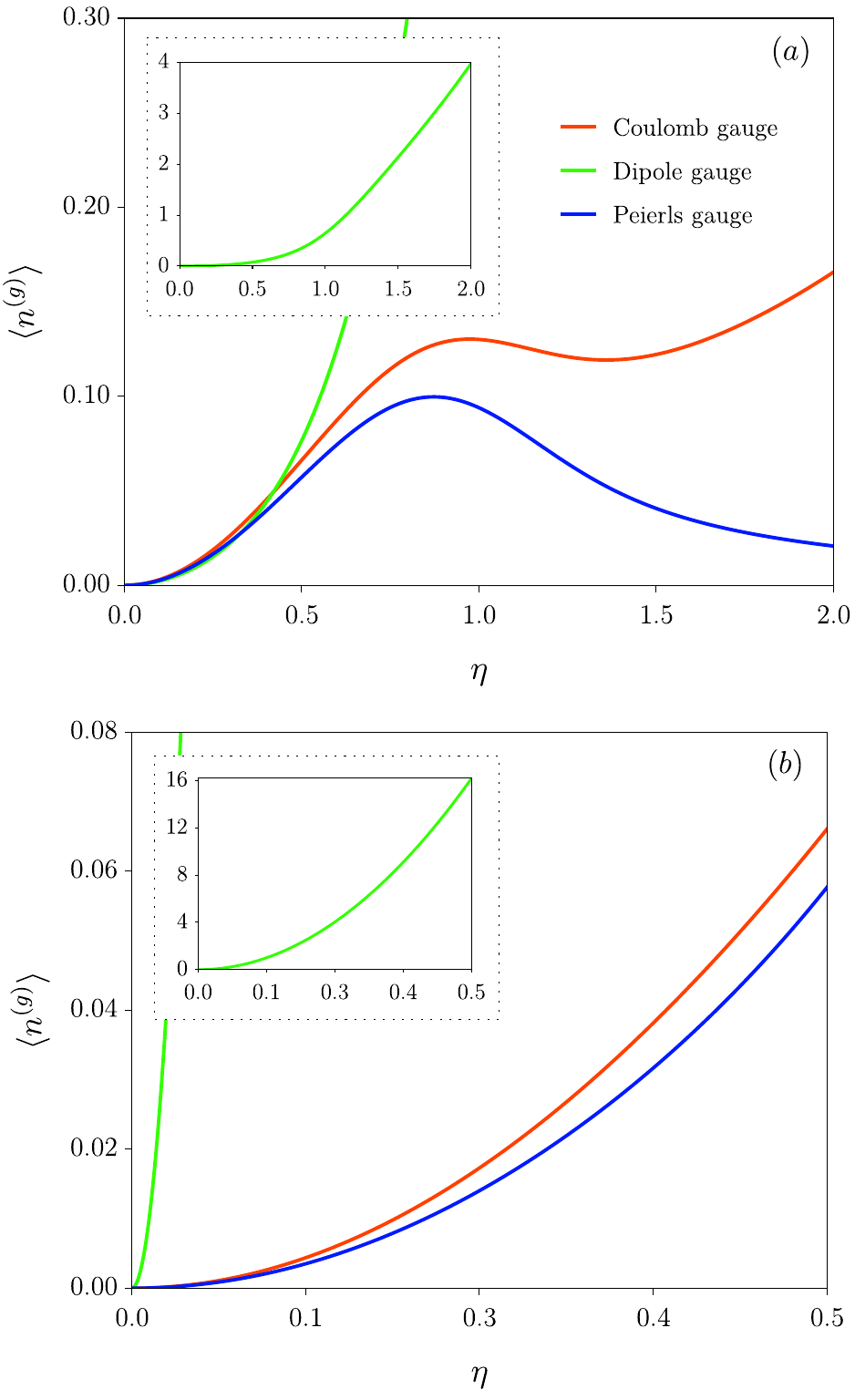}
\caption{The ground state expectation value of the photon number in each gauge $g$, $\langle n^{(g)}\rangle $, is calculated for increasing coupling $\eta$, for the regimes:~(a) $\omega=0.9\omega_{12}$ and (b) $\omega=0.9\omega_{13}$. The insets show the extended range of the dipole gauge photon number, $\langle n^{\mathrm{(dipole)}}\rangle$. }
\label{fig:photon-number}
\end{figure}

An observable which is associated with ``light'' or ``matter'' in one gauge, can be associated with mixed degrees of freedom in the next. The photon number operator is defined in a gauge $g$ by the canonical operators as $\hat{n}^{(g)}=\hat{a}^{\dag}\hat{a}|_{g}$. Each gauge introduces a unique gauge-invariant definition of the photon number operator, which can be expressed in the form 
\begin{equation}\label{eqn:photon_number}
    \hat{n}^{(g)}=\frac{1}{2\omega} \left[ (\hat{E}_{T,0}+\hat{P}_0^{(g)})^2 + \omega^2\hat{Q}_0^2\right] -\frac{1}{2}
\end{equation}
where $\hat{E}_{T,0}$ is the uniform mode of the transverse electric field. Each of these operators is associated with a different ``photon number'' observable, which provides different predictions for the ground state occupation in each gauge, as shown in Fig.~\ref{fig:photon-number}.  This is because the photon is defined relative to a different transverse field in each gauge. In the Coulomb gauge, this is the transverse electric field, $\hat{\bm{E}}_{T}=-\hat{\bm{\Pi}}|_{\mathrm{Coulomb}}$; however, in the dipole or Peierls gauge, the photon is defined in terms of a unique transverse displacement field, $\hat{\bm{D}}_{T}^{(g)}=-\hat{\bm{\Pi}}|_{g}=\hat{\bm{E}}_{T}+\hat{\bm{P}}_{T}^{(g)}$. We note in particular that the Coulomb and Peierls gauge definitions are distinct.

The ground state admits an increasing number of ``dipole gauge'' photons in Fig.~\ref{fig:photon-number}. At strong coupling, this count vastly exceeds the number of photons as defined in the Coulomb or Peierls gauge. If we work in terms of the Coulomb gauge, we can see that these photons are attributed to the measure of the electric dipole moment via $\hat{a}|_{\mathrm{dipole}}=\hat{a}|_{\mathrm{Coulomb}} - i A q \hat{D}$. Approximating $|\Psi\rangle_{\mathrm{Coulomb}}\approx|\psi\rangle_{\mathrm{m}
}\otimes|0\rangle_{\mathrm{em}}$ (for the relative scales shown in Fig.~\ref{fig:photon-number}), the growth at high coupling is due to $ \langle \hat{n}^{(\mathrm{dipole})}\rangle \approx q^2 A^2 \langle \psi | \hat{D}^2|\psi\rangle_\mathrm{Coulomb}$ in the Coulomb gauge picture. What is interesting is that the number of ``Peierls gauge'' photons is not subject to the same growth. From the point of view of the Peierls gauge's degrees of freedom, it is precisely the on-site dipole moment terms, which are excluded in this picture, which account for the ``dipole gauge'' photon occupancy at high coupling:~$ \langle \hat{n}^{\mathrm{(dipole)}}\rangle \approx q^2 A^2 \langle \psi | \hat{d}^2|\psi\rangle_{\mathrm{Peierls}}$, via $ \hat{a}|_{\mathrm{dipole}}=\hat{a}|_{\mathrm{Peierls}} - i A q \hat{d}$.

\begin{figure*}[!tp]
    \includegraphics[width=\linewidth]{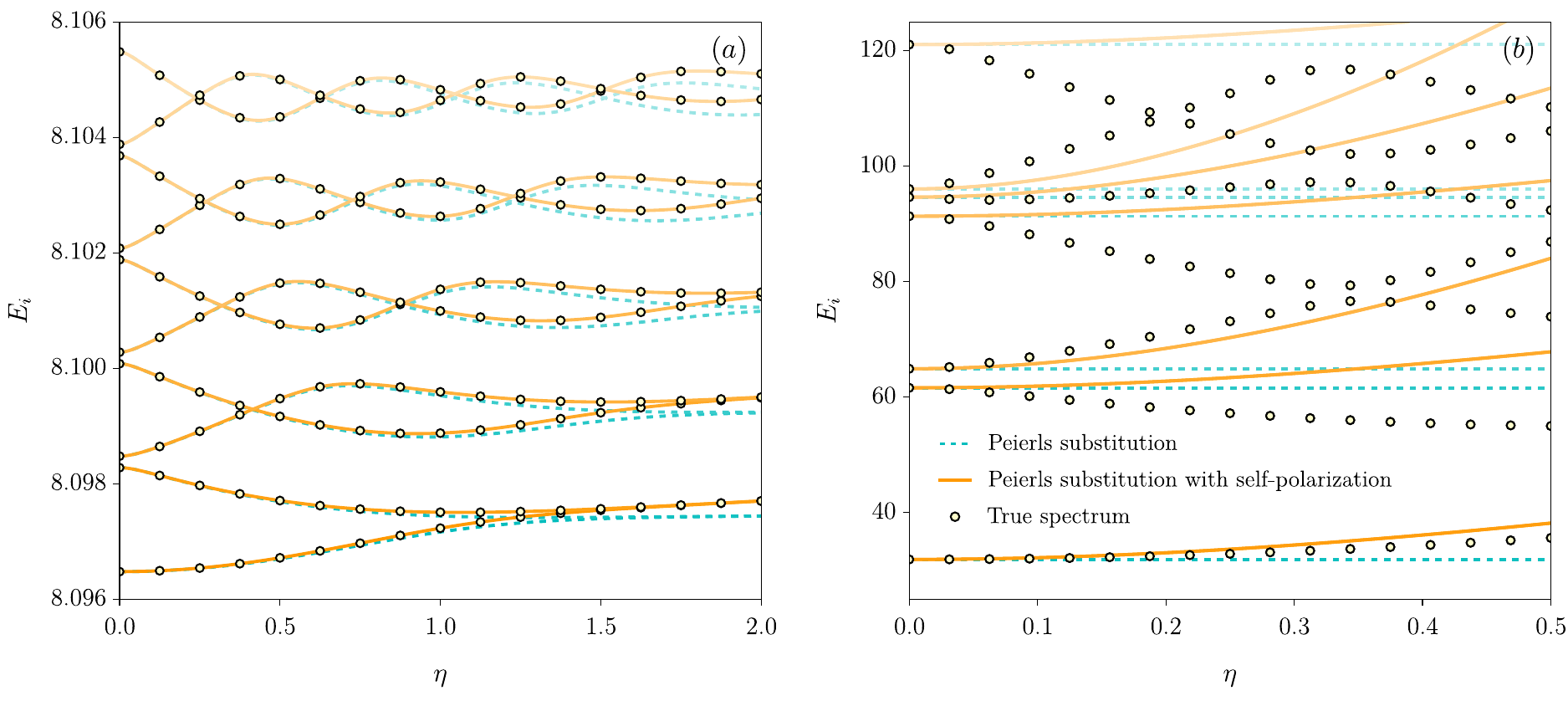}
    \caption{The energy spectrum is plotted under the Peierls substitution alone and when the self-polarization term is included, under increasing coupling $\eta$. The results are shown for the two regimes:~(a)~$\omega=0.9\omega_{12}$ (intraband coupling) and (b) $\omega=0.9\omega_{13}$ (interband coupling). The expected results from Fig.~\ref{fig:spectrum} are plotted for comparison.}
    \label{fig:pol}
\end{figure*}

\subsection{Peierls substitution and the polarization field}\label{sec:pol}

To construct a simple cavity model, the conventional approach has often been to couple an electronic system to the electromagnetic field by the Peierls substitution alone. By the Peierls substitution, we refer to the Hamiltonian 
\begin{equation}\label{eqn:qpeierls}
    \hat{H}=\sum\nolimits_{\ell\ell'} t_{\ell\ell'} e^{i q \hat{\chi}_{\ell\ell'}} \hat{c}^{\dag}_{\ell} \hat{c}_{\ell'} + \hat{H}_{\mathrm{em}},
\end{equation}
which neglects coupling to the intraband and interband polarization field terms which are found in the Peierls gauge theory, whose Hamiltonian is given in Eq.~\eqref{eq:H_peierls_quantum}.
For 1D electron models, we have discussed that the intraband dipole moments can be chosen to vanish under orbital symmetry. In this case, the Peierls substitution will capture the coupling for truncation to a single electron band, which we discuss in Sec.~\ref{sec:non-invariance}~\cite{dmytruk_gauge_2021}. The question then is whether the neglected interband dipole moments contribute to the low-energy theory, in particular via the self-polarization term, $\hat{H}_{P'^2}=\omega A^2 q^2 \hat{D}^{\prime2}$. 

For the intraband coupling, $\omega\sim\omega_{12}$, the Peierls substitution performs well at low coupling, as shown in Fig. \ref{fig:pol}(a). However, there is a growing offset with the true spectrum at stronger coupling. This offset is due to the neglect of the self-polarization term, which contains contributions from the interband dipole moments. 
We remark that the self-polarization term approximately acts as a scalar operator on the two-orbital subspace because of the symmetry of the toy model. This means that the energy differences $E_i-E_1$ and the eigenvectors are accurately captured by the Peierls substitution in the double-well model; however, we will see in Sec.~\ref{sec:non-invariance} that this does not guarantee that all observables can be captured by the Peierls substitution.
Nonetheless, the self-polarization will not possess this symmetry for a generic 1D electron system. We emphasize that, even for a single electron, the self-polarization will contribute a non-trivial correction to the spectrum and the eigenstates, which grows with stronger coupling.

We also apply the Peierls substitution to the multi-band analogue of our model. When the cavity frequency is close to resonance with the interband frequency, \mbox{$\omega\sim\omega_{13}$,} the Peierls substitution breaks down as shown in Fig.~\ref{fig:pol}(b).  We also see that the spectrum is not reproduced when the self-polarization term is included. 
To discuss this further, it is useful to know that, under a uniform field, the Peierls substitution can be unitarily transformed to a dipole gauge Hamiltonian which couples to the on-site dipole moments only, i.e.~where $\hat{D}$ is replaced by $\hat{d}$ in Eq.~\eqref{hd}~\cite{dmytruk_gauge_2021}.
In the material eigenbasis, this dipole moment operator is block-diagonal with respect to the subspace of each bound pair, 
\begin{equation}
    \hat{d}=\sum\nolimits_{\alpha=1}^{m}R_{1,\alpha}(|2\alpha-1\rangle \langle 2\alpha|+\mathrm{h.c.})
\end{equation}
with $R_{1,\alpha}=-R_{2,\alpha}$.
When these dipole transitions are far from resonance with the photonic frequency, such as when $\omega\sim (\varepsilon_{3}-\varepsilon_{1})$, they will be heavily suppressed in the dipole gauge picture. While this regime only models coupling between flat bands, it is clear in any case that all terms in Eq.~\eqref{eq:H_peierls_quantum} should be accounted for in order to model a system with multiple electron bands in the Peierls gauge.

\subsection{Gauge non-invariance under orbital truncation}\label{sec:non-invariance}

\begin{figure*}[t]
    \includegraphics[width=\linewidth]{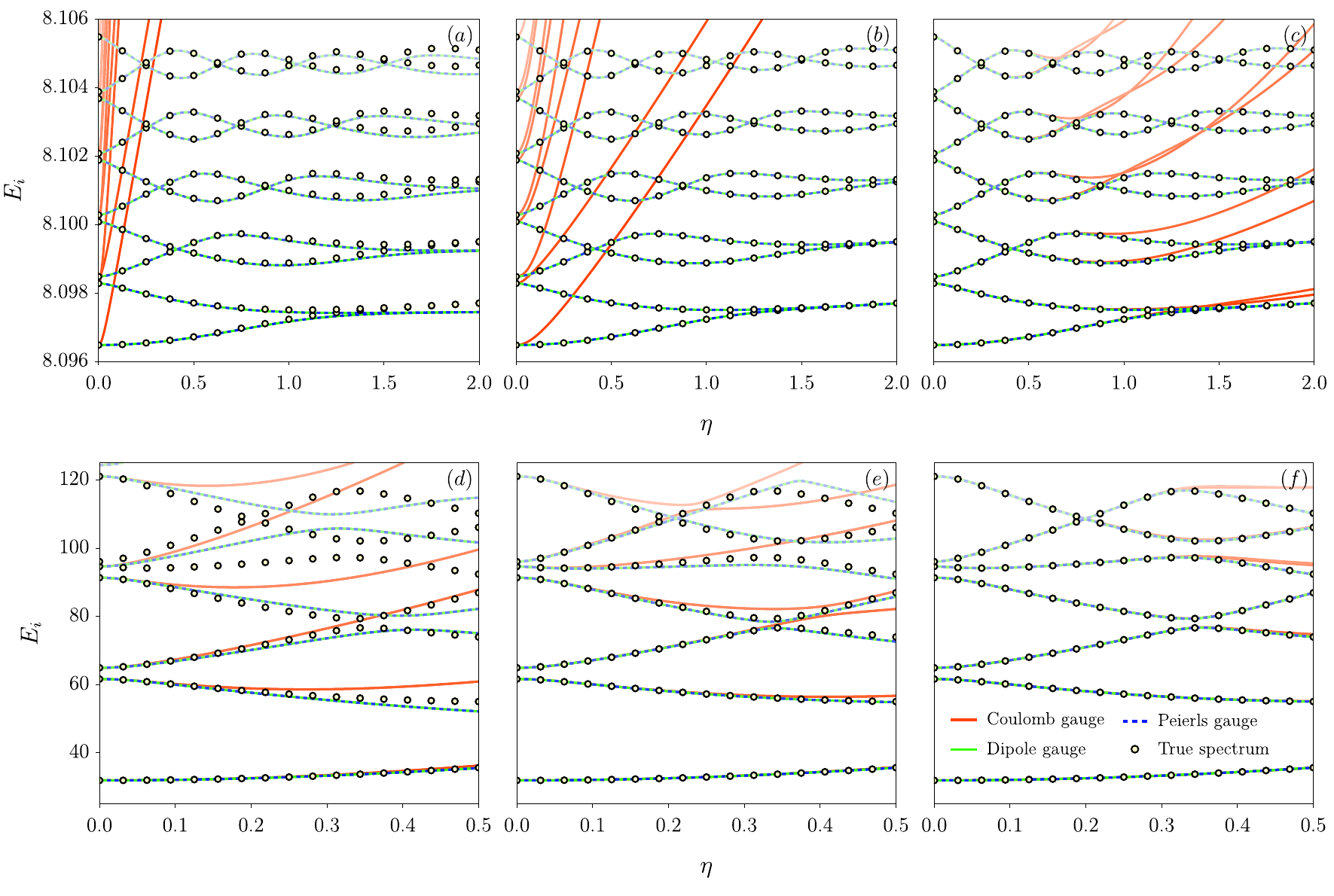}
    \caption{The spectrum is shown for material truncations which are performed on the model in the Coulomb, dipole, and Peierls gauges, under increasing coupling $\eta$. The truncations are placed as:~(a--c) $N_{\mathrm{el}}=2,4,12$ for the coupling $\omega=0.9\omega_{12}$ and (d--f) $N_{\mathrm{el}}=4,6,12$ for $\omega=0.9\omega_{13}$. The converged results from Fig.~\ref{fig:spectrum} are plotted for comparison.}
    \label{fig:trunc}
\end{figure*}

It is well known that orbital truncations will perform better in the dipole or Peierls gauges than in the Coulomb gauge in models of cavity systems~\cite{de_bernardis_breakdown_2018,li_electromagnetic_2020,ashida_cavity_2021,eyles_cauchy-schwarz_2025}.
Matrix elements connecting low- and high-energy electron orbitals are less suppressed in the Coulomb gauge compared to the dipole or Peierls gauges. This can lead to a greater dependence of the model on these transitions, even when separations in energy are large~\cite{de_bernardis_breakdown_2018}, as we can see in the relation
\begin{equation}\label{eqn:pr}
    \frac{\langle j|\hat{p}|k\rangle}{m}=i(\varepsilon_{j}-\varepsilon_{k})\langle j|\hat{x}|k\rangle,
    \end{equation}
which can be obtained via the operator identity $\hat{p}/m=i[\hat{H}_\mathrm{m},\hat{x}]$. Similar relations for periodic systems are discussed in Ref.~\cite{gu_relation_2013}.

Fig.~\ref{fig:trunc} shows the spectrum of the double-well cavity model after material truncation is performed in each gauge. The dipole and Peierls gauges demonstrate the expected advantage over the Coulomb gauge when either intraband ($\omega\sim\omega_{12}$) or interband ($\omega\sim\omega_{13}$) coupling is considered. We can see that the Peierls gauge produces the same spectrum as the dipole gauge under orbital truncation in this model. This is because the two pictures remain connected by a unitary transformation,
\begin{align}
    \hat{c}_{i\alpha}|_{\mathrm{Peierls}}
        &=e^{+iq R_{i,\alpha} \hat{A}}\, \hat{c}_{i\alpha}|_{\mathrm{dipole}},\label{eqn:equiv}
\end{align}
at any level of the truncation when subject to a uniform field, which transforms as $\hat{\Pi}_0|_{\mathrm{Peierls}}=\hat{\Pi}_0|_{\mathrm{dipole}}+\sqrt{2\omega} A \hat{d}$.

A two-level truncation reduces the light-matter coupling in our model to a Peierls substitution (or equivalent) in the dipole and Peierls gauges. For $\omega\sim\omega_{12}$, the spectrum shown in Fig.~\ref{fig:trunc}(a) demonstrates the same scalar offset shown in Fig.~\ref{fig:pol}(a), which we attribute to interband dipole moment terms that have been eliminated by the truncation. 
Including the next pair of orbitals is sufficient to reproduce the expected spectrum, as shown in Fig.~\ref{fig:trunc}(b).
We remarked in the previous section that interband corrections arise from the self-polarization term. This means that, for a well-isolated band, we can obtain an accurate model by projecting the self-polarization directly onto the one-band subspace.

\begin{figure}[t]
    \includegraphics[width=1\linewidth]{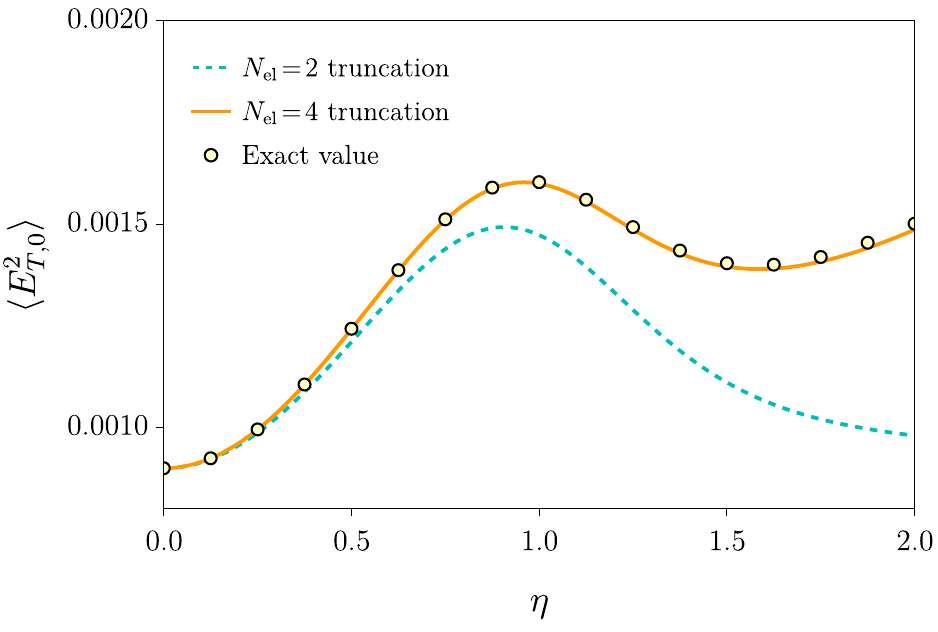}
\caption{The ground state expectation $\langle E_{T,0}^2\rangle$ is calculated under $N_{\mathrm{el}}=2,4$ orbital truncations in Peierls gauge for \mbox{$\omega\sim\omega_{12}$} under increasing $\eta$. The converged result is shown for comparison. 
}
\label{fig:tr-obs}
\end{figure}

When placing a material truncation, we should also  aim to produce accurate expectation values for any physical observables of interest. 
As an example, we consider the expectation value $\langle E_{T,0}^2\rangle$ under orbital truncation in the Peierls gauge, which we show in Fig.~\ref{fig:tr-obs}.
While the transverse electric field mode, $\hat{E}_{T,0}$, is associated with the photon field in the Coulomb gauge, this operator is sensitive to interband dipole moments in the Peierls gauge representation:~$\hat{E}_{T,0}=-\hat{\Pi}_0|_{\mathrm{Peierls}} -\hat{P}_0^{(\mathrm{Peierls})}$.
Since these terms are eliminated in the two-orbital truncation, we cannot reproduce the expected results for $\langle E_{T,0}^2\rangle$ in this case, even though the ground state is well approximated by truncation for this model. This will also be true for the Peierls substitution more generally, since the polarization field is needed to represent $\hat{E}_{T,0}$ in the Peierls gauge.

An alternative approach to the Peierls substitution can be found by truncating the position operator inside the PZW transformation. This is often taken with the aim of ``restoring'' gauge invariance between truncated Coulomb and dipole gauge models~\cite{di_stefano_resolution_2019,savasta_gauge_2021,dmytruk_gauge_2021}. For a uniform cavity field, the PZW transformation can be written in the passive view as
\begin{equation}\label{eqn:PZW}
    \hat{c}_{j}|_{\mathrm{Coulomb}}=\sum\nolimits_{k}(e^{-iq \hat{A}D})_{jk} \hat{c}_k|_{\mathrm{dipole}},
\end{equation}
with $\hat{\Pi}_0|_{\mathrm{Coulomb}}=\hat{\Pi}_0|_{\mathrm{dipole}}+\sqrt{2\omega} A \hat{D}$.
We can replace $D_{jk}$ with its two-level truncation, $D_{12}\sigma_x$, in Eq.~\eqref{eqn:PZW} to find a new unitary transformation which acts on the two-orbital dipole gauge model~\cite{savasta_gauge_2021}. 
This new transformation does not truncate the Coulomb gauge operators on the left hand side of Eq.~\eqref{eqn:PZW}, but instead coincides with the transformation to the Peierls gauge,
\begin{equation}
\hat{c}_j |_{\mathrm{Peierls}}= \sum\nolimits_{k=1,2}(e^{-iq \hat{A}D_{12}\sigma_x})_{jk} \hat{c}_k |_{\mathrm{dipole}},
\end{equation}
which generates the Peierls substitution on the two-orbital model.
This holds true for truncation to a single band $\alpha$ under a uniform field, for which the dipole and Peierls gauges are related by Eq.~\eqref{eqn:equiv} with the truncated dipole moment operator $D_{i\alpha,j\alpha}=R_{i,\alpha}\delta_{ij}$~\cite{dmytruk_gauge_2021}.

This method generates a new canonical frame within the truncated Hilbert space of the dipole gauge theory and can be extended to include multiple bands~\cite{dmytruk_gauge_2021}. These frames are distinct from the Coulomb gauge and, for truncation to a single band, we have seen that the new frame instead belongs to a truncated Peierls gauge theory. This distinction is important because the Peierls gauge does not share the same definition for ``electron'' and ``photon'' degrees of freedom with the Coulomb gauge, as we demonstrated with the photon number in Sec.~\ref{sec:photon}.
When it comes to physical observables, the representation of the associated operator is different in each of these frames, which must be accounted for to obtain the correct predictions.

\section{\label{sec:Discussion}Summary and conclusion}

We have studied a toy model of a single electron in a double-well potential which is coupled to a uniform cavity field. The potential can be tuned to describe ``bands'' of localised orbitals, which model pairs of atomic orbitals or Wannier functions on a lattice with two sites. The light-matter coupling was studied in the Coulomb, dipole, and multi-center dipole (or Peierls) gauges, with each of these gauges representing a unique separation of the composite system into ``light'' and ``matter''. 

Although the Peierls substitution can be a desirable starting point for cavity models, by itself it neglects the contribution of the polarization field terms which naturally emerge when a transformation of the Coulomb gauge to the Peierls gauge is performed. 
Moreover, we see that the Peierls substitution does not belong to the Coulomb gauge theory, but instead is defined for the same electron and photon degrees of freedom which are introduced in the Peierls gauge.

While intraband terms can often be eliminated under symmetry in 1D models, we showed that the interband dipole moments contribute to the spectrum of a single electron in the cavity at strong coupling via the self-polarization. While our  toy model  represents a peculiar case in which the low-energy spectrum is shifted only by a scalar offset, this is not the case in general. In a more general case, we would expect that the self-polarization contributes to the dressed hopping terms.  Furthermore, we found in Sec.~\ref{sec:non-invariance} that the self-polarization can also contribute to the calculation of expectation values for observables in this model.
Moreover, we showed that the Peierls substitution fails to capture the coupling between bands in our model, requiring direct coupling via dipolar polarization fields. In 2D or 3D models, the intraband dipolar coupling will also generally be needed to describe the light-matter interaction.

While gauge invariance is ensured for the full theory, projections to the low energy theory cannot preserve all possible transformations. Truncations on the material subsystem will lead to different approximations when performed in different canonical frames, which is an important consideration when modeling cavity systems~\cite{ashida_cavity_2021, de_bernardis_breakdown_2018,li_electromagnetic_2020,eyles_cauchy-schwarz_2025}.  When a finite system is coupled to a uniform field, however, the material truncation does not break gauge invariance between the Peierls and dipole gauges. These truncations outperform material truncation in the Coulomb gauge when the cavity field is coupled to intraband or interband transitions in our model. 

When constructing models for cavity quantum materials, it is worth keeping in mind that the Peierls gauge will lead to a more accurate low-energy theory than the Coulomb gauge when the  continuum fields are projected onto one or more isolated bands. However, the simpler form of the Coulomb gauge coupling may be preferable for the purpose of writing down a simple cavity model. 
Beyond this, the form of electron-electron interactions may need to be considered carefully in many-electron systems. 
These interactions can include an image charge distribution in the Coulomb gauge, which is introduced to satisfy the boundary conditions which are imposed on the total electric field~\cite{power_quantum_1982}. 
For localized subsystems, a key advantage of the multipolar framework is that the canonical photon fields coincide with the physical, local field $\hat{\bm{E}}$ outside of the material system, and the boundary conditions are therefore automatically enforced  through the photonic mode functions~\cite{power_quantum_1982,stokes_comparison_2025}.
While it is clear that the self-polarization term contributes directly to the many-body interactions the Peierls gauge, further work is needed to explore these terms for extended cavity materials.

On a final note, the gauge-relativity of the light and matter subsystems can raise ambiguity on how quantities which we associate with light or matter are defined~\cite{stokes_implications_2022}. We have shown that the Coulomb, dipole, and Peierls gauges provide different definitions for electron and photon operators, which we discussed in the context of ground state photon occupations. We can see that many definitions of electron and photon correlations, or light-matter entanglement entropy, for instance, can arise from the gauge freedom. The physical relevance of these definitions can ultimately only depend on which degrees of freedom are  accessible by experiment~\cite{stokes_implications_2022}.

\begin{acknowledgments}
A.L.~acknowledges support by the Engineering and Physical Sciences Research Council [Grant No. EP/W524347/1]. A.P.~acknowledges support from the Leverhulme Trust under the grant agreement RPG-2023-253.  The authors would like to acknowledge the assistance given by Research IT and the use of the Computational Shared Facility at The University of Manchester.
\end{acknowledgments}

\appendix

\section{\label{sec:PZW transformation}Transformation to the Peierls gauge}

\subsection{The multi-center Power-Zienau-Woolley transformation}

In the main text, we perform the transformation to the multi-center multipolar gauge by projecting the matter field onto hybridized Wannier functions. We will begin by writing the exponents of the hybridized functions as
\begin{equation}\label{eqn:gauge_choice}
    \hat{\chi}_{\ell}(\bm{x})=\int d^3x'\, \bm{g}_{T\ell}(\bm{x}',\bm{x})\cdot\hat{\bm{A}}_{T}(\bm{x}')
\end{equation}
in terms of a transverse function 
\begin{equation}\label{eq:gaugefn}
    \bm{g}_{T\ell}(\bm{x}',\bm{x})=-\int_{\bm{R}_{\ell}}^{\bm{x}} d\bm{s}\cdot \eta_{T}(\bm{s},\bm{x}')
\end{equation}
where $\eta_{T,{ij}}(\bm{x},\bm{x}')=\sum_{\nu}\phi_{\nu,i}(\bm{x}) \phi^*_{j,\nu}(\bm{x'})$ is the identity on the subspace of transverse cavity field modes. The line integral in Eq.~\eqref{eq:gaugefn} is path-dependent, which we will take as a straight line, $\bm{s}=\lambda(\bm{x}-\bm{R}_{\ell})+\bm{R}_{\ell}$, with $\lambda\in[0,1]$. The choice of the transverse function $\bm{g}_{T\ell}$, in essence, fixes the canonical frame~\cite{stokes_implications_2022}. If we instead defined each line integral with respect to the origin, we would begin the usual PZW transformation to the multipolar gauge, with all Wannier functions being rotated by the same phase, $\hat{\chi}_{\ell}=\hat{\chi}_0$.

With the phases defined in Eq.~\eqref{eqn:gauge_choice}, we begin by transforming the matter field as
\begin{equation}\label{eqn:mPZWapp}
    \hat{\psi}(\bm{x})
    =\sum\nolimits_{\ell} w_{\ell}(\bm{x}) e^{-iq \hat{\chi}_{\ell}(\bm{x})} \hat{\bar{c}}_{\ell}.
\end{equation}
The transformed site operators $\hat{\bar{c}}_{\ell}^{(\dag)}$ will preserve the fermionic canonical algebra provided the hybridized Wannier functions, $\hat{y}_{\ell}(\bm{x})=e^{-iq\hat{\chi}_{\ell}(\bm{x})} w_{\ell}(\bm{x})$, can be taken as orthogonal.
This condition will hold under the loop approximation, 
\begin{equation}\label{eqn:loop}
 \int d^3x \, w_{\ell}^*(\bm{x}) e^{{i q\hat{\Phi}_{\ell\ell'}(\bm{x})}} w_{\ell'}(\bm{x})\approx0
\end{equation}
for $\ell\neq\ell'$, which requires that the magnetic flux $\hat{\Phi}_{\ell\ell'}(\bm{x}) = \oint_{\mathcal{C}} d\bm{s}\cdot\hat{\bm{A}}_{T}(\bm{s}) $ threading a loop $\mathcal{C}$ through the points $ \bm{R}_{\ell} \to \bm{R}_{\ell'} \to \bm{x} \to \bm{R}_{\ell} $ vanishes under the overlap of each pair of Wannier functions. 

The photonic fields are related by a polarization field between the Coulomb and multipolar gauges. This field ensures that electron and photon operators commute in the new frame. We can see that the transverse magnetic potential, $\hat{\bm{A}}_{T}$, will be invariant under the transformation. However, its conjugate field will transform as
\begin{align}\label{eqn:appPiP}
     \hat{\bm{\Pi}}(\bm{x})&=\hat{\bar{\bm{\Pi}}}(\bm{x})+\hat{\bm{P}}_{T}(\bm{x})
\end{align}
where $\hat{\bm{P}}_{T}(\bm{x})=\sum_{\nu}\hat{P}_{\nu}\bm{\phi}^*_{\nu}(\bm{x})$ is a transverse field whose components satisfy the equation $[\hat{\Pi}_{\nu},\hat{\bar{c}}_{\ell}]=[\hat{P}_{\nu},\hat{\bar{c}}_{\ell}]$.
We can show that this field is given by
\begin{align}\label{eqn:multi-PT}
    \hat{\bm{P}}_{T}(\bm{x})=&\sum\nolimits_{\ell\ell'}\bm{\mathcal{P}}_{\ell\ell'}(\bm{x})\hat{c}^{\dag}_{\ell}\hat{c}_{\ell'} \\
    =&\sum\nolimits_{\ell\ell'}\bm{\mathcal{P}}_{\ell\ell'}(\bm{x}) e^{iq\hat{\chi}_{\ell\ell'}}\hat{\bar{c}}^{\dag}_{\ell}\hat{\bar{c}}_{\ell'} \notag
\end{align}
where 
\begin{equation}
\bm{\mathcal{P}}_{\ell\ell'}(\bm{x})=-q\int d^3x'\,w^*_{\ell}(\bm{x}')\bm{g}_{T\ell}(\bm{x},\bm{x}') w_{\ell'}(\bm{x}')
\end{equation}
are the matrix elements of the transverse part of a polarization-like field centered on the site $\bm{R}_{\ell}$. It follows from the loop approximation that this matrix element can also be centered on the site $\bm{R}_{\ell'}$ since $\int d^3x'\,w^*_{\ell}(\bm{x}')[\bm{g}_{T\ell}(\bm{x},\bm{x}')-\bm{g}_{T\ell'}(\bm{x},\bm{x}')] w_{\ell'}(\bm{x}')\approx0$.

The transformation removes light-matter interaction terms from the minimal coupling, up to a small magnetic correction. The full Hamiltonian is given in the multi-center multipolar gauge by
\begin{align}\label{eqn:multipolarH}
    \hat{H}=\sum\nolimits_{\ell\ell'} \hat{\tau}_{\ell\ell'}e^{iq\hat{\chi}_{\ell\ell'}} \hat{\bar{c}}^{\dag}_{\ell} \hat{\bar{c}}_{\ell'} + \hat{U}_{\mathrm{ee}}
    +\hat{\bar{H}}_{\mathrm{em}}+\hat{\bar{H}}_{P\Pi}+\hat{\bar{H}}_{P^2}
\end{align}
where $\hat{\chi}_{\ell\ell'}=\hat{\chi}_{\ell}(\bm{x})-\hat{\chi}_{\ell'}(\bm{x})$ are the Peierls factor exponents introduced in Eq.~\eqref{eq:peierls_phases}, and we have brought magnetic corrections into the light-dependent hopping amplitudes
\begin{equation}\label{eqn:tau}
   \hat{\tau}_{\ell\ell'} = \int d^3x \, w_{\ell}^{*}(\bm{x}) \!
   \left(\! \frac{(-i\bm{\nabla}-q \hat{\bm{A}}_{\ell'}(\bm{x}))^2}{2m}+ V(\bm{x})\! \right)\! w_{\ell'}(\bm{x}).
\end{equation}
Here, we can interpret the field
\begin{align}    
\hat{\bm{A}}_{\ell}(\bm{x}) &= \hat{\bm{A}}_{T}(\bm{x}) + \bm{\nabla}\hat{\chi}_{\ell} \\
&=-\int_0^1 d\lambda\,\lambda(\bm{x}-\bm{R}_{\ell}) \times \hat{\bm{B}}(\bm{R}_{\ell}+\lambda(\bm{x}-\bm{R}_{\ell}) )\notag
\end{align}
as a magnetic vector potential whose longitudinal component, $\hat{\bm{A}}_{L}(\bm{x})=\bm{\nabla}\hat{\chi}_{\ell}$, is defined uniquely with respect to each orbital $\ell$.
The final two terms in Eq.~\eqref{eqn:multipolarH} arise from the transformation of the electromagnetic field term $\hat{H}_{\mathrm{em}}$ under Eq.~\eqref{eqn:appPiP}, describing a direct multipolar interaction 
\begin{equation}
    \hat{\bar{H}}_{P\Pi}=\frac{1}{2}\int d^3x\,\hat{\bm{P}}_{T}(\bm{x})\cdot\hat{\bar{\bm{\Pi}}}(\bm{x})+ \hat{\bar{\bm{\Pi}}}(\bm{x})\cdot \hat{\bm{P}}_{T}(\bm{x})
\end{equation}
and the self-polarization term
\begin{equation}\label{eqn:self-pol}
    \hat{\bar{H}}_{P^2}=\frac{1}{2}\int d^3x\,\hat{\bm{P}}_{T}(\bm{x})^2.
\end{equation}
For many-electron systems, this self-polarization term contributes directly to the electron-electron interactions in the multipolar canonical frame.

While we do not consider these interactions in our single-electron model, it should be noted that the transformation of the Coulomb interaction term, $\hat{U}_{\mathrm{ee}}$, under Eq.~\eqref{eqn:mPZWapp} will depend on the form of the expression which was introduced with the quantization in the Coulomb gauge. As an example, we can consider the interaction term in free space, which in the Coulomb gauge is given by $\hat{U}_{\mathrm{ee}}=\sum\nolimits_{\ell\ell'mm'} U_{\ell mm'\ell'} \hat{c}_{\ell}^{\dag}\hat{c}_{m}^{\dag}\hat{c}_{m'}\hat{c}_{\ell'}$, where $U_{\ell m m' \ell'}=\frac{q^2}{8\pi}\int d^3x \int d^3x'\,w_{\ell}^*({\bm{x}})w_{m}^*({\bm{x'}})\frac{1}{|\bm{x}-\bm{x}'|}w_{m'}({\bm{x'}})w_{\ell'}({\bm{x}})$. This term is expressed as
\begin{equation}
    \hat{U}_{\mathrm{ee}}=\sum\nolimits_{\ell\ell'mm'} U_{\ell mm'\ell'} e^{iq(\hat{\chi}_{\ell\ell'}+\hat{\chi}_{mm'})} \hat{\bar{c}}_{\ell}^{\dag}\hat{\bar{c}}_{m}^{\dag}\hat{\bar{c}}_{m'}\hat{\bar{c}}_{\ell'}
\end{equation}
in the new canonical frame. However, as we discussed in Sec.~\ref{sec:theory}, the expression for $\hat{U}_{\mathrm{ee}}$ depends on the cavity boundary conditions which are imposed on the longitudinal electric field, $\hat{\bm{E}}_{L}$, in the Coulomb gauge theory, which can demand the inclusion of image charges.

\subsection{Local electric dipole approximation}

In order to recover the Peierls substitution and the dipole interaction terms, we will need to truncate the photonic modes to remove higher-order multipolar contributions. 
We introduce an electric dipole approximation (EDA) on the scale of the lattice, in which we assume the relevant modes $\nu$ satisfy
\begin{equation}\label{eqn:eda}
    w^*_{\ell}(\bm{x})\bm{\phi}_{\nu}(\bm{x})w_{\ell'}(\bm{x})\approx w^*_{\ell}(\bm{x})\bm{\phi}_{\nu}(\bm{R}_{\ell'})w_{\ell'}(\bm{x})
\end{equation}
in which it is understood that both sides of the equation vanish for any orbitals $\ell$ and $\ell'$ which are far displaced. Eq.~\eqref{eqn:eda} introduces a coarse-graining of the field mode functions as $\bm{\phi}_{\nu}(\bm{R}_{\ell})\approx\bm{\phi}_{\nu}(\bm{R}_{\ell'})$ for neighboring sites, which will be an important feature of the approximation.

For consistency, the gauge potentials are restricted to frequencies which satisfy the local EDA, $\hat{\bm{A}}_{T}(\bm{x})=\sum\nolimits_{\nu\in\Lambda} \hat{Q}_{\nu} \bm{\phi}_{\nu}(\bm{x})$ and $\hat{\bm{\bar{\Pi}}}(\bm{x})=\sum\nolimits_{\nu\in\Lambda} \hat{\bar{\Pi}}_{\nu} \bm{\phi}^{*}_{\nu}(\bm{x})$. We assume this truncation can be placed such that all physically relevant frequencies are included. 
Moreover, the truncation should be performed such that all field modes $\bm{\phi}_{\nu\notin \Lambda}(\bm{x})$ are removed from the theory, including the polarization field, which becomes $\hat{\bm{P}}_{T}(\bm{x})=\sum\nolimits_{\nu\in\Lambda} \hat{P}_{\nu} \bm{\phi}^{*}_{\nu}(\bm{x})$. This restores gauge invariance between the two theories under truncation by restricting the transformation to act on the low-frequency modes alone. Of course, this defines a unique choice of the multipolar canonical frame. This approach is often taken in the usual PZW transformation to define the dipole approximation or as a regularization to remove relativistic modes from the theory~\cite{cohen-tannoudji_photons_1997}.

The new transformation between the photon mode-truncated theories is straightforwardly given by restricting the cavity modes in the transverse function $\bm{g}_{T\ell}$ as
\begin{equation}
    \bm{g}_{T\ell}(\bm{x}',\bm{x})=-\sum\nolimits_{\nu\in\Lambda}\bm{\phi}^*_{\nu}(\bm{x'})\int_{\bm{R}_{\ell}}^{\bm{x}} d\bm{s}\cdot \bm{\phi}_{\nu}(\bm{s}).
\end{equation}
This transverse function satisfies a similar electric dipole approximation 
\begin{align}\label{eqn:2nd_approx}
    &w^*_{\ell}(\bm{x})\bm{g}_{T\ell'}(\bm{x}',\bm{x})w_{\ell'}(\bm{x}) \approx \\&
    - \sum\nolimits_{\nu\in\Lambda}  \bm{\phi}^*_{\nu}(\bm{x}')\left[w^*_{\ell}(\bm{x})(\bm{x}-\bm{R}_{\ell'})w_{\ell'}(\bm{x}) \cdot \bm{\phi}_{\nu}(\bm{R}_{\ell'})\right]\notag
\end{align}
where we have used that $w^*_{\ell}(\bm{x})\int_{\bm{R}_{\ell'}}^{\bm{x}}d\bm{s}\cdot\bm{\phi}_{\nu}(\bm{s})w_{\ell'}(\bm{x}) \approx w^*_{\ell}(\bm{x})(\bm{x}-\bm{R}_{\ell'})\bm{\phi}_{\nu}(\bm{R}_{\ell'})w_{\ell'}(\bm{x})$, under the coarse-graining of the field modes as described before.

Together, these dipole approximations simplify Eq.~\eqref{eqn:multipolarH} in two ways. Firstly, the correction terms which arise from transforming the minimal coupling terms will vanish. For pairs of localized Wannier functions, we find 
\begin{equation}
w^*_{\ell}(\bm{x})\hat{\bm{A}}_{\ell'}(\bm{x}) w_{\ell'}(\bm{x}) \approx0, 
\end{equation}
which recovers the usual tight-binding hopping amplitudes as  $\hat{\tau}_{\ell\ell'} \approx t_{\ell\ell'}.$
Secondly, the transverse polarization field is truncated to include only dipolar contributions. These are described by the modes
\begin{equation}
 \hat{P}_{\nu} =\sum\nolimits_{\ell\ell'}\mathcal{P}_{\nu,\ell\ell}\,e^{iq\hat{\chi}_{\ell\ell'}}\hat{\bar{c}}^{\dag}_{\ell}\hat{\bar{c}}_{\ell'} 
 \end{equation}
whose polarization field matrix elements
\begin{equation}
    \mathcal{P}_{\nu,\ell\ell'}\approx q \bm{\phi}_{\nu}(\bm{R}_{\ell'})\cdot\bm{D}'_{\ell\ell'}
\end{equation}
are given in terms of the site-centered dipole moments $\bm{D}'_{\ell\ell'}=\int d^3x\,w^*_{\ell}(\bm{x})(\bm{x}-\bm{R}_{\ell'})w_{\ell'}(\bm{x})$.
Together, these results lead to the multi-center dipole (or Peierls) gauge Hamiltonian which is given in Eq.~\eqref{eq:H_peierls_quantum} in the main text.

\bibliography{References}

\end{document}